 \documentclass[reprint,showpacs,preprintnumbers,amsmath,amssymb,aps,prb,longbibliography,lengthcheck]{revtex4-1}
 \usepackage{graphicx}% Include figure files
 \usepackage{dcolumn}% Align table columns on decimal point
 \usepackage{bm}% bold math
 \usepackage{hyperref}% add hypertext capabilities
\begin{document}
 
 \preprint{R. E. F\'elix-Medina \textit{et al }}
 
 \title{Magnetism of small V clusters embedded in a Cu fcc matrix: an \textit{ab initio} study}% Force line breaks with \\
 
 \author{R. E. F\'elix-Medina$^{1}$, M. A. Leyva-Lucero$^{1}$, R. A. Guirado-L\'opez$^{2}$, S. Meza-Aguilar$^{1}$}
 \affiliation{ 
 $^{1}$ Escuela de Ciencias F\'{\i}sico-Matem\'aticas, 
 Universidad Aut\'onoma de Sinaloa, 
 Blvd. de las Am\'ericas y Universitarios, 
 Ciudad Universitaria, Culiac\'an Sinaloa,
 CP 80010, M\'exico. \\
 $^{2}$ Instituto de F\'{\i}sica, ``Manuel Sandoval Vallarta'', 
 Universidad Aut\'onoma de San Luis Potos\'{\i}, \'Alvaro Obreg\'on
 64, 78000 San Luis Potos\'{\i}, M\'exico.}

 \date{January 23, 2010}% It is always \today, today,
 
 \begin{abstract}
 We present extensive first principles density functional theory (DFT)
calculations dedicated to analyze the magnetic and electronic properties of
small V$_{n}$ clusters (n=1,2,3,4,5,6) embedded in a Cu fcc matrix. We consider
different cluster structures such as: i) a single V impurity, ii) several V$_{2}$
dimers having different interatomic distance and varying local atomic
environment,  iii) V$_{3}$ and iv) V$_{4}$ clusters for which we assume compact as well
as 2- and 1-dimensional atomic configurations and finally,  in the case of the
v) V$_{5}$ and vi) V$_{6}$ structures  we consider a square pyramid and a square
bipyramid together with linear arrays, respectively. In all cases, the V atoms
are embedded as substitutional impurities in the Cu network. In general, and
as in the free standing case, we have found that the V clusters tend to form
compact atomic arrays within the cooper matrix. Our calculated non
spin-polarized density of states at the V sites shows a complex peaked
structure around the Fermi level that strongly changes as a function of both
the interatomic distance and local atomic environment, a result that
anticipates a non trivial magnetic behavior. In fact, our DFT calculations
reveal, in each one of our clusters systems, the existence of different
magnetic solutions (ferromagnetic, ferrimagnetic, and antiferromagnetic) with
very small energy differences among them, a result that could lead to the
existence of complex finite-temperature magnetic properties. Finally, we
compare our results with recent experimental measurements.
 \end{abstract}
 
 %\keywords{Suggested}%Use showkeys class option if keyword
 \maketitle
  
 \section{\label{sec:Introduction} Introduction}
 
 The discovery of magnetism in V metal fine particles \cite {Akoh} with sizes 
 and temperatures ranging from 90 to 300 \AA\ and from 4.2 to 100 K, 
 respectively, served as a starting point for a large number of theoretical 
 and experimental studies of magnetic properties in various types of V 
 low-dimensional systems 
 \cite{Turek,Bryk,Robles,Bouarab,Cocula,Lacina,Raul,BergmanPRB,Bergman,Qiu,Valla}. In particular, 
 bimetallic V-based materials such as bulk 
 alloys \cite {Qiu}, thin films \cite {Valla}, and deposited 
 clusters \cite{Bergman,Weber} have attracted considerable attention because of 
 their expected novel optical properties, catalytic activity, and magnetic 
 behavior.
  
 In general the interaction between two components in magnetic bimetallic 
 nanostructures introduces a mutual influence on neighboring atoms and leads 
 to the unique properties already reported for these kinds of materials. 
 Furthermore, it is also clear that their measured average magnetic behavior 
 is not necessarily given by the average properties of their corresponding 
 isolated constituents, and that the observed phenomena are expected to 
 strongly depend on the precise details of the local geometrical and chemical 
 environment.
  
 As is well known, V in the solid state is nonmagnetic; however, it has been 
 clearly established that it could exhibit a magnetic moment under certain 
 conditions like loss of coordination, when its atomic volume is increased, or 
 when it is found in the presence of 3d ferromagnetic materials, due to its
 large paramagnetic susceptibility. Interestingly, in the last years it has
 been also demonstrated that V can become magnetic when it is associated with
 other nonmagnetic elements such as copper \cite{Weber} or gold 
 \cite{Galanakis}. The previous findings are highly relevant since, on the one 
 hand, having two nonmagnetic elements that when combined give raise to 
 magnetic order opens new possibilities to improve the magnetic properties of 
 materials and, on the other hand, from the theoretical point of view it 
 offers a unique opportunity to analyze the role of the structure and 
 dimensionality on the spontaneous appearance of local magnetic moments in 
 metallic nanostructured materials.
  
 Huttel {\it et al} \cite{Huttel} have found experimentally the existence of a magnetic moment in 
 V atoms embedded in a Cu matrix. In particular, they have analyzed
 samples having a Cu$_{96.7}$V$_{3.3}$ composition by means of x-ray absorption
  spectroscopy and x-ray magnetic circular dichroism experiments. The samples
  were prepared by a co-evaporation process of both V and Cu over a
  Cu substrate and the as-prepared Cu-V samples were capped with a few
  monolayers of Cu in order to avoid oxidation of the surface alloy.
  Interestingly the authors especulate that, depending on the deposition
  temperature, different V species seems to be formed (such as isolated V
  impurities and V particles of different sizes) and that the magnetic
  properties of the alloy strongly depend on the presence and concentration of
  these species. They also showed that the V magnetic moment depends on the
  formation temperature of the samples and found that the alloys prepared at
  lower temperatures are more magnetic. However, despite all the previous
  extensive experimental characterization and reported trends, there is still no
  clear information concerning the precise micro-structural features present in
  the alloys, neither a clear explanation for the temperature dependence of
  the average magnetization observed in the samples.  
  
  From the point of view of theory, it is important to mention that the possible
  existence of structural changes, disordered growth, and clustering of the V
  species in the samples makes more difficult the theoretical analysis of the
  measured data. However, we believe that the Cu-V alloys synthesized by Huttel
  and co-workers can be approximately modeled by an infinite Cu matrix with
  V atoms and clusters embedded in it. In addition, we expect that the
  influence of the Cu matrix on the V atoms will be mainly due to the
  hybridization of the V $d$-electrons with the $s$-levels of the 
 neighboring
  Cu atoms and, in this sense, a local analysis in which the behavior of the
  spin moments at the V sites is studied for different geometrical and chemical
  environments should be very important and needs to be performed in order to
  shed some more light in to the possible origin of the measured magnetization data.
  
  In this work, we present a systematic first principles theoretical study
  of the magnetic properties of model V-dilute fcc Cu-V alloys by using the
  density functional theory (DFT) approach together with the pseudopotential
  approximation for the electron-ion interaction. All the calculations are
  performed with the Plane Wave Self Consistent Field (PWscf) 
 code \cite{Baroni}. We put special emphasis on: i) the electronic and magnetic properties of isolated atoms 
 as well as of small clusters embedded in a fcc Cu matrix, and ii) the 
 possible existence of multiple magnetic solutions in the systems, since if 
 small energy differences are found among them magnetic fluctuations could be 
 present in the samples.
 
 The remainder of the paper is organized as follows. In the next section, the 
 theoretical model used for the calculations is briefly explained. In Sec. \ref{sec:Results} 
 results for the atomic spin magnetic moments, average spin magnetic moments, 
 and differences in total energies (DTEs) for a variety of V$_{N}$ clusters 
 embedded in a Cu fcc matrix with $N\le 4$ atoms, and some special 
 cases of $N=5$ (square pyramide) and $N=6$ (square bipyramide), as well as 
 infinite linear chain are presented and discussed. For each system we 
 have studied the nonmagnetic (NM), ferromagnetic (FM), and antiferromagnetic 
 (AF) solutions, i.e, we have used these magnetic configurations as  
 inputs in our calculations. Finally, Sec. \ref{sec:Conclusions} summarizes the main conclusions.
  
  \begin{figure}
  \begin{center}
 \resizebox{0.47\textwidth}{!}{\includegraphics{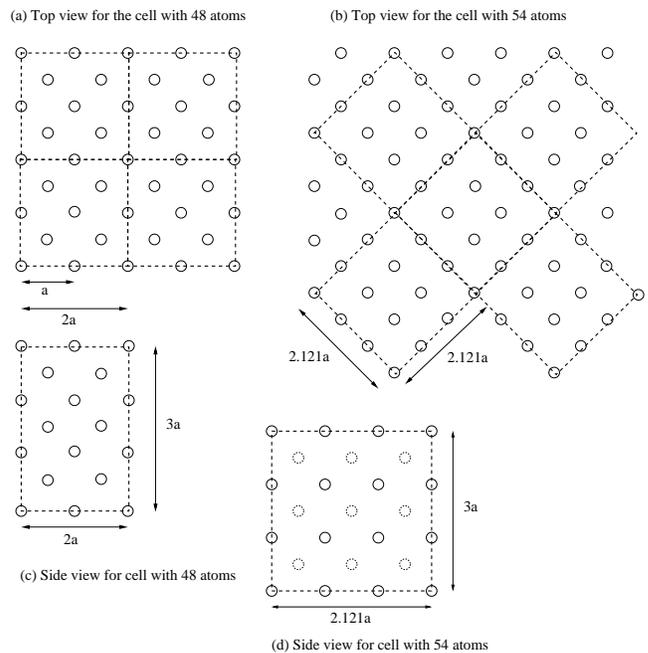}} 
  \end{center}
  \caption{\label{fig:CrystalCell} Schema of the crystal cell used.
  (a) Top view for the crystal cell with 48 atoms, 
  (b) top view for the cell with 54 atoms,
  (c) and (d) are side view for the crystal cell with 48 and 54 atoms, respectively.
  The dotted line is the crystal cell and the $a$ is the Cu lattice parameter (3.61 \AA\ ). 
  The empty circles represent the Cu or V atoms.}
  \end{figure}

  \begin{figure*} 
  \begin{center} 
 \resizebox{1.0\textwidth}{!}{\includegraphics{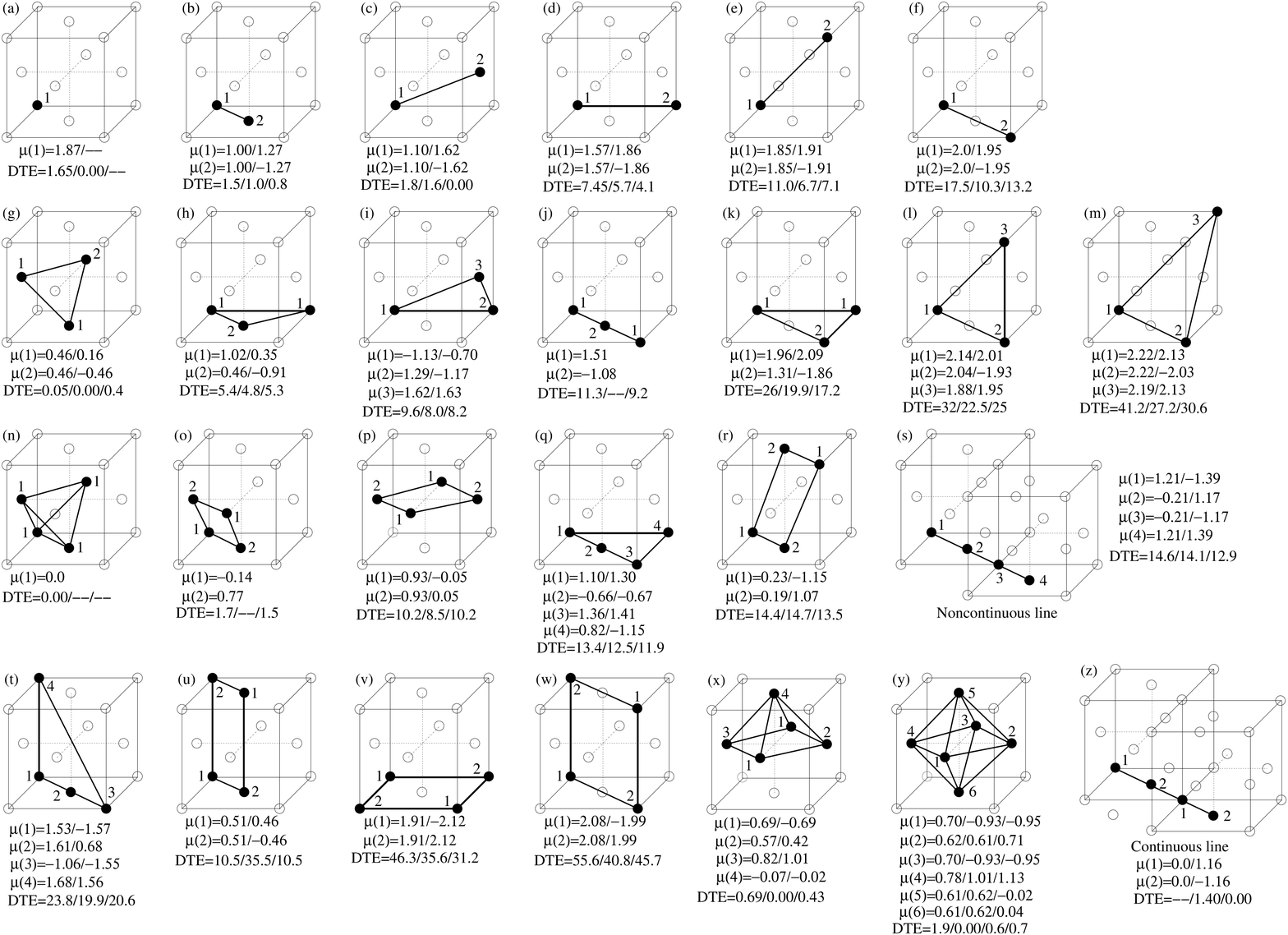}}
 % \resizebox{1.0\textwidth}{!}{\includegraphics{Clusters_Structures_raul.eps}} 
  \end{center} 
  \caption{\label{fig:geometry} 
 Illustration of the lattice structure of V$_N$ clusters embedded as 
 substitutional impurities in bulk Cu (fcc structure). The cluster atoms are 
 represented by filled circles and the Cu host atoms by open circles. The 
 numbers label the nonequivalent atomic sites $i$. 
 $\mu (i)=\mu_{FM}/\mu_{AF}$ refers to the magnetic moments (in $\mu_{B}$) 
 for the 
 ferromagnetic (FM) and antiferromagnetic (AF) solutions, respectively. 
 DTE=DTE$_{NM}$/DTE$_{FM}$/DTE$_{AF}$ 
 refers to the differences in total energies (DTEs) 
 (in $meV/Atom$) for the nonmagnetic (NM), FM, and AF solutions, respectively. 
 In the special case of the 
 square bipyramide [Fig. \ref{fig:geometry} (y)] we include two kinds of antiferromagnetic 
 solutions, namely, AF1 and AF2, whose initial magnetic configurations are,
 respectively, 
 $(1 \uparrow,2 \downarrow,3 \uparrow,4 \downarrow,5 \uparrow,6 \uparrow)$ and 
 $(1\uparrow,2 \downarrow,3 \uparrow,4 \downarrow,5 \uparrow,6 \downarrow)$.
 In this case the DTEs are presented as follows: 
 DTE=DTE$_{NM}$/DTE$_{FM}$/DTE$_{AF1}$/DTE$_{AF2}$. 
 The ground state energies (denoted by $0.00$) refer to the minimum energies 
 out of V$_N$ clusters with the same $N$. 
  } 
  \end{figure*}
  
  \section{\label{sec:Theoretical} Theoretical Model}
  
  The PWscf\cite{Baroni} code is based on the DFT. The pseudopotentials used here were 
 generated using Vanderbilt code \cite{Vanderbilt} with scalar relativistic 
 calculations and non linear core corrections. The calculations, which were 
 restricted to fixed atomic positions (relaxation was not allowed), were 
 performed with the generalized gradient approximation of 
 Perdew-Burke-Ernzerhof (GGA-PBE) \cite{GGA-PBE} exchange correlation functional. 
  The calculations were performed  using a fcc structure with Cu lattice parameter
 (3.61 \AA\ ) and the cell containing 
 48 (54) atoms (see Fig. \ref{fig:CrystalCell}), 
 ordered as follows: six planes with 8(9) atoms per plane for the
 impurity, dimer, trimer and tetramer (pyramide and bipyramide), and the V atoms
 are located at the ideal positions of the fcc crystal.  
  The Monkhorst-Pack scheme was used to define the $\vec k$ 
  points and the calculations for each cell were performed with a grid of
  4$\times$4$\times$4 mesh in $\vec k$-space. We have done some tests 
 with a grid of 5$\times$5$\times$5 and $6\times6\times6$ mesh, and the
  values are only 4$\%$ less than those presented here.
  A cutoff energy of 35 $Ry$ was used for the plane waves expansion of the
  pseudowave functions (560 $Ry$ for the 
  charge density and potential). 
 
 Given that our 
 calculations were restricted to fixed atomic positions, some words about the 
 relaxation effects in this kind of systems seem to be in order. On the one 
 hand, Ramanathan {\it et al} \cite{Ramanathan}, 
 who have studied the magnetism of a V overlayer on Nb(001), have found 
 that the magnetism is drastically reduced when the relaxation effects are 
 considered. On the other hand, Weber {\it et al} \cite{Weber}, who have
 studied the magnetism of V clusters supported on Cu(001), have analysed the 
 effect of relaxation qualitatively, and they have concluded that relaxation 
 does not yield any significant change in the magnetic moments. 
 Following that, we performed calculations for the V impurity and dimer 
 using a cell with 54 atoms, first without considering relaxation and then 
considering partial relaxation (only nearest neighbors could reach the 
equilibrium positions) and we have obtained that the differences in 
percentage terms between non-relaxation and partial relaxation calculations 
 are $\Delta E < 4\%  $ ;  $\Delta d = \sim 1.0\%$ and $\Delta \mu = \sim 2\%$ 
 for the difference of total energies, difference of distances, and 
 difference of magnetic moments, respectively.
 
 \section{\label{sec:Results} Results and discussion}
 
 \subsection{Geometrical construction}
 
 The V clusters geometric arrangements (as well as the lattice structure 
 around them) that we have studied are schematically illustrated in 
 Fig. \ref{fig:geometry}. The V embedded clusters are located as 
 substitutional impurities in the Cu fcc matrix. We
 present the atomic magnetic moments as $\mu (i) =\mu_{FM}/\mu_{AF}$ in $\mu_B$ 
 for the FM and
 AF solutions, respectively. We also present the difference in total energies as
 $DTE=DTE_{NM}/DTE_{FM}/DTE_{AF}$ 
 (in $meV/atom$)
 for the NM, FM, and AF solutions, respectively. In the special case of the 
 square bipyramide [Fig. \ref{fig:geometry} (y)] we present the difference in total energies as
 following 
 $DTE=DTE_{NM}$
 \noindent
 $/DTE_{FM}/DTE_{AF1}/DTE_{AF2}/$ where $AF1$ and $AF2$ stand, 
 respectively, for the next initial magnetic configurations: 
 $(1 \uparrow,2 \downarrow,3 \uparrow,4 \downarrow,5 \uparrow,6 \uparrow)$ and 
 $(1\uparrow,2 \downarrow,3 \uparrow,4 \downarrow,5 \uparrow,6 \downarrow)$.
 The ground states energies
 (denoted by $0.00$) refer to the minimum energies out of V$_N$ clusters with 
 fixed $N$. In all the cases below, $a =3.61$ \AA\ is the Cu lattice parameter. 
 Fig. \ref{fig:geometry} (a) represents the V impurity. The V 
 dimers are also in the first row: 
  (b), (c), (d), (e), and (f). The respective interatomic distances are:
  $a/ \sqrt {2}$, $a\sqrt{3/2}$, $a$, $a \sqrt{3}$, and $a \sqrt{2}$. The V
  trimers are in the second row. (g) represents the smallest equilateral 
 triangle
  whose sides measure $a/ \sqrt {2}$; (h) an isosceles right angle 
 triangle 
  whose hypotenuse and legs measure, respectively, $a$ and  $a/ \sqrt {2}$; 
 (i) a right angle triangle whose hypotenuse measures $a\sqrt{3/2}$, and whose legs measures
  $a/ \sqrt {2}$  and $a$; (j) a linear trimer whose interatomic
  distances are equal to $a/ \sqrt {2}$; (k) an isosceles right angle triangle 
  whose hypotenuse and legs measure, respectively, $a\sqrt {2}$ and $a$; 
 (l) a  right angle triangle 
  whose hypotenuse measures $a \sqrt{3}$, and whose legs measure $a$ 
  and $a\sqrt {2}$; and
  (m) an equilateral triangle whose sides measure $a\sqrt
  {2}$. The V tetramers are in the third and fourth rows. (n) represents a regular tetrahedron whose interatomic distances are $a/ \sqrt {2}$;
  (o) a regular rhombus whose sides measure $a/ \sqrt {2}$; 
  (p) a square whose sides measure $a/ \sqrt {2}$; (q) a
  triangular tetramer with two bonds of length equal to $a/ \sqrt {2}$ and 
 two bonds with length equal to $a$; (r) a parallelogram whose sides measure 
 $a\sqrt{3/2}$ and $a/ \sqrt {2}$; (s) a continuous
  linear tetramer whose interatomic distances are equal to $a/ \sqrt {2}$; 
  (t) a triangular tetramer with two bonds of length equal to $a/ \sqrt {2}$, 
 one bond of length equal to $a$, and one bond of length equal to 
 $a \sqrt {3}$; (u) a
  rectangle with sides equal to $a/ \sqrt {2}$ and $a$; (v) a square 
 with side equal to $a$; 
  (w) a rectangle with sides equal to $a\sqrt {2}$
  and $a$; the special cases are (x) a square pyramid 
 whose base measures $a/\sqrt {2} \times a/\sqrt {2}$ and whose sides are all 
 equilateral triangles of edge-length parameter equal to $a/ \sqrt {2}$;
 (y) a square bipyramid built from two pyramids 
 connected base to base; and finally (z) a continuous linear chain whose 
 interatomic distances are equal to $a/ \sqrt {2}$ 
 
 Fig. \ref{fig:geometry} displays the well-known richness and complexity in the magnetic 
 behaviour characterizing the 3$d$ transition metal low-dimensional systems. 
 There we can see four kinds of magnetic orders: ferromagnetic, ferrimagnetic, 
 antiferromagnetic, and nonmagnetic, which depend in a very complicated way on
 the interatomic distances, the geometrical structures, and the number of 
 atoms in the clusters. 
 
 \subsection{Electronic behavior}
  
\begin{figure}[t!]
  \begin{center}
  \resizebox{0.48\textwidth}{!}{\includegraphics{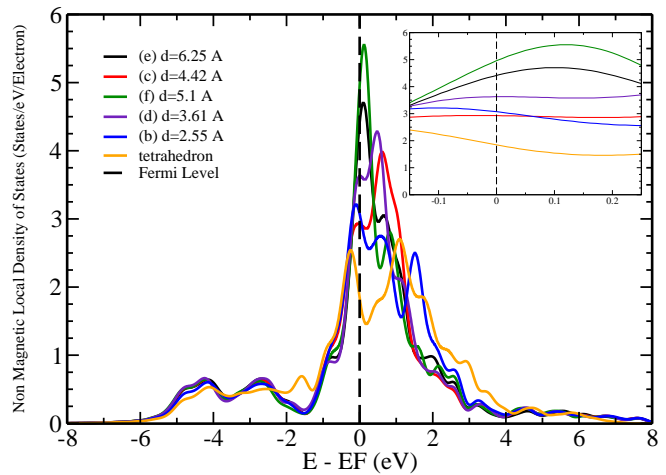}}
  \end{center} 
  \caption{\label{fig:LDOS} 
 The nonmagnetic local density of states (NMLDOS) for V atoms (only $d$
  component). The blue, red, violet, black, and green lines
  represent, respectively, the NLDOS for the dimers illustrated in 
 Fig. \ref{fig:geometry} (b), (c), (d), (e), (f). The orange line represents the NLDOS for the 
 tetrahedron tetramer [Fig. \ref{fig:geometry} (n)]. The dashed line represents the Fermi level.
  The inset shows a magnified view of the LDOS near the Fermi level.}
  \end{figure}

 Let us first briefly consider the electronic properties derived of our 
 calculations on V$_N$ clusters embedded in a Cu fcc matrix. Fig. \ref{fig:LDOS}
 illustrates the 
 $d$-component of the nonmagnetic local densities of states (NMLDOS) for all 
 the V dimers and the tetrahedron. The blue, red, violet, black, and 
 green lines represent, respectively, the NMLDOS for the V dimers illustrated 
 in Fig. \ref{fig:geometry} (b), (c), (d), (e), and (f); while the orange line represents the
 NMDOS for the tetrahedron [Fig. \ref{fig:geometry} (n)]. To study the electronic properties of
 these systems we use the Stoner criterion. 
 This is given by the following relation:
  
 \begin{displaymath}
 S=I_{S}N_{NM}(E_{F})
 \label{LDOS}
 \end{displaymath}
  
 \noindent being $S$ the Stoner factor, $I_{S}$ the Stoner parameter and 
 $N_{NM}(E_{F})$ the value of the NMLDOS at the Fermi level. The $I_{S}$ for 
 V is equal to 0.3536 eV \cite{Janak}. The Stoner criterion states that the 
 nonmagnetic state will be unstable to magnetism if {\it S} $>1$, while 
 the nonmagnetic state will be more favourable if {\it S} $\leq 1$.
 
 The behaviour of the NMLDOS in Fig. \ref{fig:LDOS} is similar to that 
 reported by Liu, Khana, and Jena \cite{Liu}, who studied the magnetism of 
 free small V clusters, in the sense that near the Fermi energy the NMLDOS 
 splits into two peaks with the Fermi energy centered in the valley. This 
 splitting is remarkably pronounced in the case of the tetrahedron. We can see 
 that for all the V dimers the value of the NMLDOS at the Fermi level are high 
 enough (greater than $1/I_S=2.8280$) to get a Stoner factor {\it S} $>1$; 
 while for the tetrahedron tetramer the 
 NMLDOS at the Fermi level is not (it is much less than $1/I_S=2.8280$). So 
 from the Stoner criterion alone we can predict magnetism for all the V dimer 
 and absence of magnetism for the tetrahedron tetramer. The greatest value of 
 the NMLDOS at Fermi level corresponds to the dimer whose interatomic distance 
 is $a\sqrt{2}=5.1$ \AA\ 
 (see \ref{fig:geometry} (f) and green line in Fig. \ref{fig:LDOS}), which, according to 
 the values of the magnetic moments shown in Fig. \ref{fig:geometry}, is that with the greatest 
 magnetic moment out of the V dimers. Additionally, from the NMLDOS we can 
 anticipate a nonlinear behaviour of the magnetic moments of V dimers as a 
 function of the interatomic distances.

 \subsection{Magnetic properties}
 
 Let us now consider the magnetic behaviour of each cluster as a function of 
 its size. For the V impurity, the magnetic moment is 1.87 $\mu_{B}$, which is 
 less than that of the V adsorbated atom on Cu(001) surface reported by 
 Weber {\it et al} \cite{Weber} (3.03 $\mu_{B}$) and 
 greater than that of the bulk impurity reported by Bl\"ugel {\it et al} 
 \cite{Blugel2} (1.10 $\mu_{B}$).
 
\begin{figure}[t]
 \begin{center}
 \resizebox{0.48\textwidth}{!}{\includegraphics{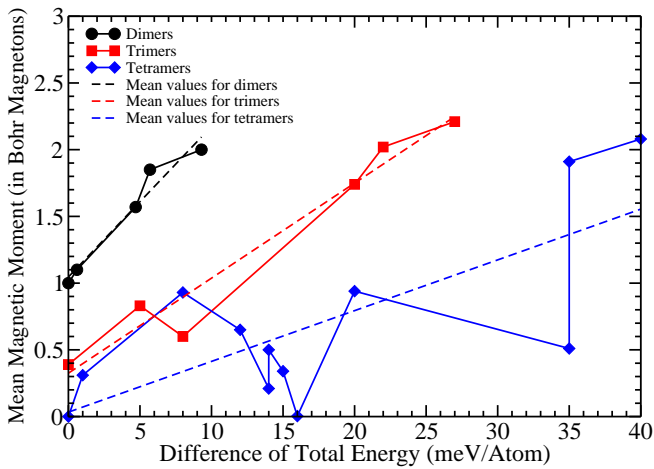}}
%\resizebox{0.53\textwidth}{!}{\includegraphics{GroundStateMeanMagneticMoments.jpg}}
 \end{center} 
 \caption{\label{fig:mmm} 
 Mean magnetic moments as a function of the difference of the total energy. 
 The black, red and blue lines are related to V dimers, trimers, and 
 tetramers, respectively. The dashed lines represent the corresponding linear 
 regressions.} 
 \end{figure}

 For the V dimers, we can see in Fig. \ref{fig:geometry} (b)-(f) that the magnetic states are 
 energetically ordered in such a way that  
 $E_{NM} > E_{FM} > E_{AF}$ ($E_{NM} > E_{AF} > E_{FM}$) provided that the 
 interatomic distance $d \leq a$ ($d > a$). So, in the first (second) case the 
 V atoms that form the dimers couple 
 antiferromagnetically (ferromagnetically). In any case, for the V dimers, 
 it is energetically easier to show some kind of magnetic coupling 
 than to become nonmagnetic. Interestingly, the V dimer ground state does not 
 correspond to the V dimer with the minimum interatomic distance, namely 
 $a/\sqrt{2}$, but to that with the third shortest one: $a \sqrt{3/2}$ 
 [see Fig. \ref{fig:geometry} (b) and (c)]. The magnetic moment per V atom for the ground state 
 V dimer is equal to $\pm$ 1.62 $\mu_{B}$ which is less than adsorbate dimer 
 magnetic moment reported by Weber {\it  et al.} \cite{Weber} 
 (2.85 $\mu_{B}$), but greater than the bulk dimer magnetic moment reported by 
 Bl\"ugel {\it et al.} \cite{Blugel2} (0.00 $\mu_{B}$). The first three 
 excited states correspond to the V dimer with $d=a/\sqrt{2}$. The 
 DTEs for these states are equal to 0.8, 1.0, and 1.51 $meV/atom$, 
 respectively, for the antiferromagnetic ($\mu = \pm 1.27 \mu_{B}$), 
 ferromagnetic ($\mu = 1.00 \mu_{B}$), and nonmagnetic solutions. So, from the 
 ground state, the most probable transition is that in which the V dimer 
 reduces both its interatomic distance and the magnitude of its magnetic 
 moment, and keeps the same kind of magnetic coupling (antiferromagnetic in 
 this case). Note that it is more probable that from the ground state the V 
 dimer reduces its interatomic distance than it changes its coupling from 
 antiferromagnetic to ferromagnetic, since for the first (second) transition 
 it needs to increase its energy in 0.8 $meV/atom$ (1.6 $meV/atom)$. For the 
 minimum interatomic 
 distance, a transition from antiferromagnetic to ferromagnetic coupling can 
 occur if the V dimer increases its energy in 0.2 $meV/atom$. This transition 
 implies a
 reduction of the magnitude of the magnetic moment from 1.27 $\mu_{B}$ to 
 1.00 $\mu_{B}$. In general, the magnitudes of the magnetic moments for the
 ferromagnetic solutions are less than those for the antiferromagntic ones (the
 only exception is that of the V dimer with $d=a \sqrt{2}$ in Fig. \ref{fig:geometry} (f)). 
 Also, in general, the magnitude of the magnetic moments tend to increase 
 (although not monotonously) as the interatomic distance increases. However, in 
 Fig. \ref{fig:mmm}, we can observe that the magnitude of the magnetic moments increase
 as the DTEs increase in a strictly monotonous way. 
 Note also that although the 
 interatomic distance for the V dimer represented in Fig. \ref{fig:geometry} (e) is greater 
 than that for the V dimer represented in Fig. \ref{fig:geometry} (f) its magnetic moment is 
 smaller 
 (1.85 $\mu_{B}$ vs. 2 $\mu_{B}$, and 1.91 $\mu_{B}$ vs 1.95 $\mu_{B}$, for 
 the FM and AF solutions, respectively) maybe due to the presence of a Cu atom 
 between the V atoms. Finally, to compare our calculations 
  with the measurements made by Huttel {\it  et al.} \cite{Huttel}, we 
 note first that for the V dimer the concentration 
 [Cu$_{46/48}$V$_{2/48}$ (Cu$_{0.96}$V$_{0.04}$)] coincides approximately with that reported 
 by Huttel {\it  et al.} \cite{Huttel} (Cu$_{0.967}$V$_{0.033}$)
  who reported a mean magnetic moment in V atoms of approximately 
 1.00 $\mu_{B}$, which nicely agrees with the V dimer first state (FM).    
 
 In general, the trimers couple ferromagnetically (the only exceptions are
 the linear chain in Fig. \ref{fig:geometry} (j) and the isosceles triangle in Fig. \ref{fig:geometry} (k)). The
 order of the magnetic states is not so clearly defined in terms of the
 geometric characteristics of the clusters as it was in the case of the 
 V dimers. As in the case of the V dimers, for the V trimers, it is 
 energetically easier to show some kind of magnetic coupling than to become 
 nonmagnetic. The V trimer ground state corresponds to the more compact 
 equilateral triangle in the Fig. \ref{fig:geometry} (g), whose interatomic distances are 
 equal to $a/\sqrt{2}$. Practically, this ground state has 2-fold degeneracy
 since the difference in the DTE for two states of lowest energy is only 
 0.05 $meV/atom$. So, we can say that there the ferromagnetic and nonmagnetic 
 states coexist. For the ferromagnetic solution, the magnetic moment per V atom 
 is equal to 0.46 $\mu_{B}$. If we think of this trimer as a superior stage
 in the evolution of the dimer in Fig. \ref{fig:geometry} (b) as the number of atoms increase, 
 we note, as expected, a diminishing in the magnetic moment as a consequence
 of the increment in the coordination number. For the V trimers in Figs. \ref{fig:geometry} 
 (g)-(j), it is easier to change their magnetic coupling than to change their
 geometric configuration. The opposite occurs for the V trimers in Figs. \ref{fig:geometry}
 (k)-(m), for example, if the V trimer in Fig. \ref{fig:geometry} (k) is in its ferromagnetic 
 state, it costs less energy to change its geometric configuration to that 
 shown in Fig. \ref{fig:geometry} (l) than to become nonmagnetic while keeping the initial 
 geometric configuration.  Finally, we note that for the linear trimer 
 (Fig. \ref{fig:geometry} (i)) the magnetic moments are $1.51$ $\mu_{B}$, 
 $-1.08$ $\mu_{B}$ and $1.51$ $\mu_{B}$, these are lower than the moments 
 found by Weber {\it et al.} \cite{Weber} in the adchain of 3 V atoms on 
 Cu(001) fcc.
  
 It is not easy for the V tetramers either to give general rules about their 
 magnetic behaviour in terms of the energies of their states. The ground state 
 corresponds to the triangular pyramide. This state has 3-fold denegeracy, 
 i.e., there coexists the nonmagnetic, ferromagnetic, and antiferromagnetic 
 states. In any case, the magnetic moments per V atom is equal to zero 
 [see Fig. \ref{fig:geometry} (n)]. This does not mean, however, that the V tetramer represents 
 a critical size of V clusters from which the magnetism vanishes since as 
 it will be seen below, V clusters with five and six atoms give nonzero 
 magnetic moments.  So, according to our calculations the V magnetic moments 
 fall to zero more slowly than those reported by 
  Bl\"ugel {\it et al.} \cite{Blugel2}. In general, we note that for small 
 V clusters embedded in a Cu fcc matrix the most stable structures trend to 
 form compact arrangements: the smallest equilateral trimer and the 
 triangular pyramide. The exception is the V dimer, whose ground state occurs 
 for the dimer with the third shortest interatomic distance. For the 
 tetramers, the first excited state corresponds to the V rhombus in 
 Fig. \ref{fig:geometry} (o). This state corresponds to the ferromagnetic solution but in 
 reality the V atoms couple
 ferrimagnetically and has a mean magnetic moment of 0.31 $\mu_{B}$. 
 The linear chain [Fig. \ref{fig:geometry} (s)] shows a ferrimagnetic order. As expected, its
 magnetic moments are smaller than those corresponding to the V dimer and the
 V linear chain trimer. 
 
 Beyond $N=4$, we have considered three other systems, namely, square 
 pyramide ($N=5$) and square bipyramide ($N=6$), and continuous linear chain. 
 The ground 
 state for the square pyramide corresponds to the ferromagnetic solution with
 a mean magnetic moment per V atom equal to $0.70 \mu_{B}$. For the square
 bipyramide we have considered four solutions, namely, nonmagnetic, 
 ferromagnetic, antiferromagnetic one (AF1), and antiferromagnetic two (AF2). 
 The input for AF1 and AF2 are, respectively:  
 $(1 \uparrow,2 \downarrow,3 \uparrow,4 \downarrow,5 \uparrow,6 \uparrow)$ and 
 $(1\uparrow,2 \downarrow,3 \uparrow,4 \downarrow,5 \uparrow,6 \downarrow)$ 
 where the numbers label the V atomic
 sites and the arrows the directions of the input spin magnetic moments. The
 ground state corresponds to the ferromagnetic solution with a mean magnetic 
 moment per V atom equal to 0.39 $\mu_{B}$. 
 Practically, the AF1 and AF2 solutions are degenerate. 
 For continuous linear chain, we have studied both ferromagnetic and antiferromagnetic solutions. 
 The ground state is the antiferromagnetic solution with magnetic 
 moments at V sites equal to $\pm 1.16 \mu_{B}$. 
 If consider the evolution of 
 the linear chain from the dimer to the infinite linear chain keeping the same 
 interatomic distance $a/\sqrt{2}$ [see Fig. \ref{fig:geometry} (b), (j), (s), and (z)], we 
 note the V atomos always couple antiferromagnetically. Besides, the V atoms
 located at the extreme sites of the chain trend to the limit of the 
 infinite chain more slowly (1.27, 1.51, 1.39, 1.16 $\mu_{B}$) than the V 
 atoms located at inner sites of the chain (1.08, 1.17, 1.16 $\mu_{B}$). 
 We think that for the square pyramide, square bipyramide, and infinite 
linear chain an increase of the degeneracy can be obtained, because it 
is possible to find other magnetic states (antiferromagnetic by blocks, for 
example). For $N>7$, we also think that it is possible to find magnetic 
degenerate states with magnetic moments per V atom eventually diminishing to 
zero (for $N$ large enough).
 
  In the Fig. \ref{fig:mmm} 
  we illustrate the mean magnetic moments as a function as the DTE, only for the FM state.  
  For V dimers, the mean magnetic moments increase as the DTE increases. 
  We can say that for room temperatures, it will be possible to find higher magnetic moments. 
  For V trimers and tetramers, the behaviour is more complicated, even so we can observe that the mean values 
  show a positive slope. 
  
 \section{\label{sec:Conclusions} Conclusions}

 We conclude our work as follows: 
 $i)$ we have studied the magnetic behaviour of small V clusters 
 ($N=1,2,3$ and 4 atoms) and some other special geometries for $N=5$ (square pyramide), 
 and $N=6$ (square bipyramide), as well as an infinite linear chain embedded in a Cu fcc matrix using an 
 \textit{ab initio} code; $ii)$ we have reproduced theoretically an 
 experimental result by Huttel\cite{Huttel} for a V concentration of approximately
 Cu$_{0.96}$V$_{0.04}$; $iii)$ for the ferromagnetic solutions, the magnetic 
 moments per V atom decrease as 
 the cluster size increases; this magnetic behaviour is in agreement with 
 previous theoretical studies \cite{Weber,Sosa}; $iv)$ the magnetic moments 
 per V atom have a strong dependence of the geometric structure \cite{Weber}; 
 $v)$ the mean magnetic moment increases with the difference of total energy; 
 $vi)$ compact structures have lower energy (they have lower mean magnetic 
 moment too), an exception being the V dimer which prefers the third 
 interatomic distance; whereas extended structures have both higher energy and mean 
 magnetic moments; 
 $vii)$ for each one of our considerd cluster systems we have found 
the existence of different magnetic solutions (nonmagnetic, ferromagnetic,
and antiferromagnetic) with very small energy differences among them, a result
that we believe could lead to the existence of complex finite-temperature
magnetic properties.

 \begin{acknowledgments}
We acknowledge the support by the Universidad Aut\'onoma de Sinaloa and SEP
through Project PIFI 2005-25-06 UAS-ECFM. 
R.A. G.-L. acknowledges the financial support from PROMEP-CA and CONACyT
through grant 50650.
 \end{acknowledgments}

 \end{document}